\documentclass[twocolumn,showpacs,preprintnumbers,amsmath,amssymb]{revtex4}
\usepackage{amsmath}

\begin{document}

\tolerance=5000
\def\pp{{\, \mid \hskip -1.5mm =}}
\def\cL{{\cal L}}
\def\be{\begin{equation}}
\def\ee{\end{equation}}
\def\bea{\begin{eqnarray}}
\def\eea{\end{eqnarray}}
\def\tr{{\rm tr}\, }
\def\nn{\nonumber \\}
\def\e{{\rm e}}
\def\D{{D \hskip -3mm /\,}}

\def\SEH{S_{\rm EH}}
\def\SGH{S_{\rm GH}}
\def\AdS5{{{\rm AdS}_5}}
\def\S4{{{\rm S}_4}}
\def\gfv{{g_{(5)}}}
\def\gfr{{g_{(4)}}}
\def\SC{{S_{\rm C}}}
\def\RH{{R_{\rm H}}}

\title{Is brane cosmology predictable?}

\author{Shin'ichi Nojiri\footnote{Electronic mail: 
snojiri@yukawa.kyoto-u.ac.jp, nojiri@cc.nda.ac.jp}}
\address{Department of Applied Physics,
National Defence Academy,
Hashirimizu Yokosuka 239-8686, JAPAN}

\author{Sergei D. Odintsov\footnote{Electronic mail: 
odintsov@ieec.fcr.es Also at TSPU, Tomsk, Russia and IFT, UNESP, Sao Paulo (temporary)}}
\address{Instituci\`o Catalana de Recerca i Estudis 
Avan\c{c}ats (ICREA) and 
Institut d'Estudis Espacials de Catalunya (IEEC), 
Edifici Nexus, Gran Capit\`a 2-4, 08034 Barcelona, SPAIN }

\begin{abstract}

The creation of the inflationary brane universe in 5d bulk Einstein 
and Einstein-Gauss-Bonnet gravity is considered. 
We demonstrate that emerging universe is ambigious due to arbitrary 
function dependence 
of the junction conditions (or freedom in the choice of boundary terms).
We argue that some fundamental physical principle (which may be related 
with AdS/CFT correspondence) is necessary in order to fix the 4d geometry in unique way. 

\end{abstract}


\maketitle

Despite the number of the efforts the construction of the theory of all fundamental
interactions is still very far from the end. At the moment (super)string/M-theory 
remains to be the most promising candidate of unified theory.
String/M-theory lives in higher dimensional space from which the early 4d 
universe naturally emerges in brane-world approach \cite{RS1}.
Brane cosmology (for recent  review and list of references, see \cite{maartens, lidsey}) is somehow different 
from usual 4d cosmology due to the presence of extra terms having 
the higher dimensional origin. 
Nevertheless, there appeared the arguments that early brane universe may be quite a realistic possibility.

It remains not completely clear for all followers of brane-worlds that 
brane cosmology is not predictable without some additional physical principle.
The junction conditions (or cosmological equations) are ambigious (up to the arbitrary function) and do not 
define the four-dimensional geometry in the unique way. In the present letter 
we demonstrate this explicitly for two models: 5d Einstein 
and 5d Einstein-Gauss-Bonnet (EGB) gravity. As a brane universe the inflationary
(deSitter) space is considered. It is shown that creation of dS instanton  
always occurs just 
by corresponding choice of junction conditions (or by the choice of surface terms). 
The predictibility of brane cosmology may be achieved only when fundamental physical principle (for instance,
AdS/CFT correspondence) is applied to fix the surface action. 

Let us start from 5d Einstein gravity where two (identified with each other)  bulk spacetimes are glued 
 at the 4d surface (boundary). 
The starting action $S$  is the sum of the Einstein-Hilbert action $\SEH$, the Gibbons-Hawking 
surface term $\SGH$ \cite{GH}  and the surface action $S_1$:
\bea
\label{Stotal}
S&=&\SEH + \SGH + 2 S_1 \\
\label{SEHi}
\SEH&=&{1 \over 16\pi G}\int d^5 x \sqrt{g}\left(R_{(5)} 
+ {12 \over l^2}\right) \\
\label{GHi}
\SGH&=&{1 \over 8\pi G}\int d^4 x \sqrt{h}\nabla_\mu n^\mu \ .
\eea
Here $n_\mu$ is the unit vector perpendicular to the surface, with direction of $n_\mu$  to be inside the bulk. 
The induced metric is denoted as $h_{\mu\nu}=g_{\mu\nu} - n_\mu n_\nu$. 
The Euclidean signature is used but changing $g\to -g$ and $h\to -h$ the Lorentzian signature results are obtained.

By the variation over the metric, the surface equation follows:
\be
\label{GK1}
K_{\mu\nu} - K h_{\mu\nu} = 8\pi G t_{\mu\nu}\ .
\ee
Here, the extrinsic curvature $K_{\mu\nu}$ is defined to be $K_{\mu\nu}=\nabla_\mu n_\nu$, 
$K\equiv h^{\mu\nu}K_{\mu\nu}$ and the surface stress tensor $t_{\mu\nu}$ is
$t^{\mu\nu}\equiv \left(2/\sqrt{h}\right)\delta S_1/\delta h_{\mu\nu}$ .
Eq.(\ref{GK1}) is nothing but the Israel junction condition \cite{Israel} (see also \cite{CR}). 
 It is customary to start from bulk AdS space:  
\be
\label{GK4}
ds^2 = dz^2 + l^2 \sinh^2 \frac{z}{l} d\Omega_{(4)}^2\ .
\ee
Here $\Omega_{(4)}^2$ is the metric of the 4d sphere with unit radius, $n_z=n^z=-1$, 
$n_\mu=n^\mu=0$ ($\mu\neq z$) and $\nabla_\mu n_\nu = \left(1/l\right)\coth\left(z/l\right) h_{\mu\nu}$.
Let the surface action is
\be
\label{GK6}
S_1=\int d^4 x \sqrt{h}  \left\{ -\frac{3}{8\pi Gl} + L_m\right\}\ ,
\ee
where
$t_{\mu\nu}=- \left(3/8\pi Gl\right)h_{\mu\nu}+ t_{m\, \mu\nu}$ and $t_{m\, \mu\nu}$ denotes the contribution 
from $L_m$. Assuming there is a boundary (sphere) at $z=z_0$ Eq.(\ref{GK1}) looks like 
$\coth\left(z_0/l\right)=1 + \cdots $. Here $\cdots$ expresses the possible contribution from 
$L_m$ ($t_{m\,mu\nu}$). One can define the brane radius $R$ as $R\equiv l\sinh\left(z_0/l\right)$ to rewrite 
the brane equation in the form $0=\left(1/R\right)\sqrt{1 + R^2/l^2} - 1/l + \cdots$ .

As $S_1$ one may consider  more complicated action
\be
\label{GK8}
S_1=\int d^4 x \sqrt{h}  \left\{ -\frac{3}{8\pi Gl'} +  \frac{\alpha}{8\pi G}R_{(4)} + \cdots
\right\}\ .
\ee
Here $R_{(4)}$ is  4d scalar curvature induced on the surface and  
the terms containing the higher powers of the curvature invariants are not written explicitly. 
This kind of the surface action has been introduced in order to cancel the divergence of the bulk action 
(to make AdS/CFT correspondence to be well-defined). 
Then 
\bea
\label{GK9}
t_{\mu\nu}&=&\left( -\frac{3}{8\pi Gl'} +  \frac{\alpha}{8\pi G}R_{(4)} + \cdots \right)h_{\mu\nu} 
 - \frac{2\alpha}{8\pi G}R_{(4)\,\mu\nu} \nn
&& + \cdots \ ,
\eea
and the junction condition (\ref{GK1}) gives the brane equation
\be
\label{GK10}
 -\frac{3}{R}\sqrt{1 + \frac{R^2}{l^2}} = - \frac{3}{l'} + \frac{6\alpha}{R^2} + \cdots \ .
\ee
We now expand l.h.s. in (\ref{GK10}) with respect to $l^2/R^2$ as
$-\left(3/R\right)\sqrt{1 + R^2/l^2} = - 3/l - 3l/2R^2 + \cdots $.
The following choice $l'=l$, $\alpha=- l/4$, $\cdots$ can convert (\ref{GK10}) into an identity. 
Conversely, if $S_1$ is not included, it may be induced on the boundary. 
The coefficients correspond to those in the surface counterterms in \cite{EJM}. 

If  one changes the surface action $S_1$, the junction condition (\ref{GK1}) is also changed. 
For example, we consider the action like
\be
\label{GG1}
S_1=\frac{1}{16\pi G}\int d^4x \sqrt{h}f\left(R_{(4)}\right)\ .
\ee
with an arbitrary function $f$. 
For the same four-dimensional sphere one gets
\be
\label{GG3}
t_{\mu\nu}=\frac{1}{8\pi G}h_{\mu\nu}\left\{\frac{1}{2}f\left(\frac{12}{R^2}\right) - \frac{3}{R^2}f'\left(
\frac{12}{R^2}\right)\right\}\ ,
\ee
while the junction condition is
\be
\label{GG4}
 -\frac{3}{R}\sqrt{1 + \frac{R^2}{l^2}} = \frac{1}{2}f\left(\frac{12}{R^2}\right) - \frac{3}{R^2}f'\left(
\frac{12}{R^2}\right)\ ,
\ee
It depends from the arbitrary function $f$.
The situation is strange because the brane cosmological equation could be changed rather arbitrary by the 
choice of $f$. The question about brane universe predictability appears! 
Indeed, imagine we define a function $g\left(R_{(4)}\right)$ as
\bea
\label{GGbb}
g\left(R_{(4)}\right)&\equiv& \frac{1}{2}f\left(\frac{12}{R^2}\right) - \frac{3}{R^2}f'\left(
\frac{12}{R^2}\right) \nn
&=& \frac{1}{2}f\left(R_{(4)}\right) - \frac{R_{(4)}}{4}f'\left(R_{(4)}\right)\ ,
\eea
then
$f\left(R_{(4)}\right)=-4R_{(4)}^2\int^{R_{(4)}} dx \frac{g(x)}{x^3}$ .
Thus, an arbitrary junction condition in a form as $-\left(3/R\right)\sqrt{1 + R^2/l^2} 
= g\left(R_{(4)}\right)$ could be realized simply by the choice 
$f\left(R_{(4)}\right)=-4R_{(4)}^2\int^{R_{(4)}} dx g(x)/x^3$. 
For example, with the choice
\be
\label{GGdd}
f\left(R_{(4)}\right)=12R_{(4)}^2\int^{R_{(4)}} \frac{dx}{x^3}\sqrt{\frac{x}{12} + \frac{1}{l^2}}\ ,
\ee
which leads to $g\left(R_{(4)}\right)=-3\sqrt{R_{(4)}/12 + 1/l^2}=-\left(3/R\right)\sqrt{1 + R^2/l^2}$, 
the junction condition (\ref{GG4}) is identically satisfied. 
On the other hand, since $-\left(3/R\right)\sqrt{1 + R^2/l^2}<-3/l$, if  $g\left(R_{(4)}\right)>-3/l$, 
the brane equation (\ref{GG4}) has no solution. This shows that brane-world  is lacking the physical 
principle to predict in unique way the surface term and, hence, the emerging brane cosmology. 
 One may prescribe the AdS/CFT correspondence to define the boundary 
action\cite{EJM} where curvature surface terms correspond to usual counterterms 
in dual QFT. Of course, this may be satisfactory only in case the string theory is the fundamental theory 
(which is not yet clear). As surface action is defined there, even if brane equation has no solution one may 
take into account the other effects (brane conformal anomaly) to get the inflationary brane 
cosmology induced by quantum effects as is done in \cite{zerbini}.     
Note also that for the original Randall-Sundrum model\cite{RS1} the situation is simpler.
Indeed, branes are flat there and only constant term enters the surface action which is ambigious only due to 
possible rescaling of the brane tension. 

Another theory which received a lot of attention from brane-world point of view is  5d EGB gravity. 
It may look even more attractive than Einstein gravity because its field equations are also 
of the second order but one more parameter (GB coupling constant) presents here. 
Various aspects of EGB gravity from brane cosmology to AdS black holes and holography 
were investigated in refs.\cite{GB,GB1,GB2} (see also refs.therein). 
Let us demonstrate that the same ambiguity of junction condition occurs in EGB theory.
First, we consider the junction condition for the higher derivative (HD) gravity using surface 
counterterms found in \cite{NOjhep,GB}.  The action of 5d $R^2$-gravity is:
\bea
\label{ai}
S_{R^2}&=&\int d^5 x \sqrt{-g}\Bigl\{a R_{(5)}^2 + b R_{(5)\,\mu\nu} R_{(5)}^{\mu\nu} \nn
&& + c  R_{(5)\,\mu\nu\xi\sigma} R_{(5)}^{\mu\nu\xi\sigma} + {1 \over \kappa^2} R_{(5)} - \Lambda \Bigr\}\ .
\eea
By introducing auxiliary fields $A$, $B_{\mu\nu}$, and $C_{\mu\nu\rho\sigma}$, one can rewrite the action  
(\ref{ai}) in the following form:
\bea
\label{iia}
\lefteqn{S=\int d^5 x \sqrt{-g}\Bigl\{a \left(2 A R_{(5)} - A^2\right)} \nn 
&& + b \left(2 B_{\mu\nu} R_{(5)}^{\mu\nu} - B_{\mu\nu} B^{\mu\nu}\right) \\
&& + c \left(2 C_{\mu\nu\xi\sigma} R_{(5)}^{\mu\nu\xi\sigma} - C_{\mu\nu\xi\sigma} C^{\mu\nu\xi\sigma}\right)
+ {1 \over \kappa^2} R_{(5)} - \Lambda \Bigr\}\ .\nonumber
\eea
Here $\kappa^2=16\pi G$. 
Using the equation of the motion
\be
\label{iib}
A=R_{(5)}\ ,\quad B_{\mu\nu}= R_{(5)\,\mu\nu}\ ,\quad 
C_{\mu\nu\rho\sigma}= R_{(5)\,\mu\nu\rho\sigma}\ , 
\ee
the action (\ref{iia}) is equivalent to (\ref{ai}). 
Let us impose a Dirichlet type boundary condition, which is consistent with (\ref{iib}), 
$A=\left. R_{(5)}\right|_{\rm on\ the\ boundary}$, 
$B_{\mu\nu}=\left. R_{(5)\,\mu\nu}\right|_{\rm on\ the\ boundary}$, 
and $C_{\mu\nu\rho\sigma}=\left. R_{(5)\,\mu\nu\rho\sigma} \right|_{\rm on\ the\ boundary}$ and 
$\delta A = \delta B_{\mu\nu} = \delta C_{\mu\nu\rho\sigma}=0$ on the boundary. However, 
the conditions for $B_{\mu\nu}$ and $C_{\mu\nu\rho\sigma}$ are, in general, inconsistent. 
For example, even if $\delta B_{\mu\nu}=0$, $\delta B_{\mu}^{\ \nu}=\delta g^{\nu\rho}B_{\mu\rho}\neq 0$. 
Then one can impose boundary conditions on the scalar quantities:
\bea
\label{bc1}
&& A= B_\mu^{\ \mu}= C^{\mu\ \nu}_{\ \mu\ \nu}= R_{(5)}\ ,\nn
&& n^\mu n^\nu B_{\mu\nu} = n^\mu n^\nu C_{\mu\rho\nu}^{\ \ \ \rho}= n^\mu n^\nu R_{(5)\,\mu\nu} \ .
\eea
and 
\bea
\label{bc2}
&& \delta A=\delta\left( B_\mu^{\ \mu}\right)
=\delta \left( C^{\mu\ \nu}_{\ \mu\ \nu}\right)
=\delta\left( n^\mu n^\nu B_{\mu\nu}\right) \nn
&& =\delta\left( n^\mu n^\nu C_{\mu\rho\nu}^{\ \ \ \rho}\right)=0\ .
\eea
We now add to the action the surface terms $S_b$ corresponding to Gibbons-Hawking term and $S_1$  
\bea
\label{Iiv}
&& S_b = \int d^4 x \sqrt{- g}\Bigl[
4 a \nabla_\mu n^\mu A + 2 b\left(n_\mu n_\nu \nabla_\sigma n^\sigma \right. \nn
&& \left. + \nabla_\mu n_\nu \right) B^{\mu\nu} + 8 c n_\mu n_\nu \nabla_\tau n_\sigma C^{\mu\tau\nu\sigma} 
+ {2 \over \kappa^2}\nabla_\mu n^\mu \Bigr] \ .\nonumber
\eea
The above boundary action  makes the variational 
procedure of HD theory to be well-defined\cite{NOjhep}.
The Gauss-Bonnet combination for $R^2$ terms corresponds to $a=c$, $b=-4c$.
It is assumed the warped bulk metric is $ds^2= dz^2 + \e^{2\tilde A(z)}\tilde g_{ij}dx^i dx^j$
and there is a boundary surface (brane) at $z=z_0$. 
Then if $t_{\mu\nu}$  
 is proportional to $h_{\mu\nu}$ as $t_{\mu\nu}=t h_{\mu\nu}$,
 the (function dependent) junction condition 
is given by 
\bea
\label{GBc2}
&& 0= 8 c \left(2R_{(5),z} - 4 R_{(5),zz,z}  - \tilde R^{\ \ \ \ i}_{(5)\,i\ ,z} + C^{\ \ \ i}_{ziz\ ,z}\right) \nn
&& + \left\{ 8c \left(6A - 24 B_{zz} + 7 R_{(5)\,zz} - R_{(5)\,i}^{\ \ \ \ i}\right) 
+ \frac{24}{\kappa^2}\right\} \tilde A_{,z} \nn
&& - 4t\ .
\eea

For GB theory, the length parameter $l$ of the bulk AdS space is given by 
$1/l^2=\left(1/4c\kappa^2\right)\left(1 \pm \sqrt{1 + 2c\Lambda \kappa^4/3}\right)$. Then Eq.(\ref{GBc2}) reduces to
\be
\label{GBc4}
0=6\left(-\frac{12c}{l^2} + \frac{1}{\kappa^2}\right) \tilde A_{,z} + t\ .
\ee
For the choice of bulk metric as in (\ref{GK4}), Eq.(\ref{GBc4}) becomes:
\be
\label{GBc5}
 -\left(-\frac{12c\kappa^2}{l^2} + 1\right)\frac{3}{R}\sqrt{1 + \frac{R^2}{l^2}} = \frac{2}{\kappa^2}t\ .
\ee
Especially  with the choice of $S_1$ (\ref{GG1}) we have
\be
\label{GBc6}
 -\left(-\frac{12c\kappa^2}{l^2} + 1\right)\frac{3}{R}\sqrt{1 + \frac{R^2}{l^2}} 
= \frac{1}{2}f\left(\frac{12}{R^2}\right) - \frac{3}{R^2}f'\left(\frac{12}{R^2}\right)\ .
\ee
Hence, the junction condition could be modified rather arbitrary by the  
choice of $f$. 
The EGB brane cosmology is not predictable as well as in Einstein theory.
In \cite{NOjhep,GB}, $f\left(R_{(4)}\right)$ has been chosen to be a constant:
\be
\label{JHEP}
f\left(R_{(4)}\right)=6\eta\ .
\ee
Then the junction condition (\ref{GBc6}) which defines the creation of 
 dS universe has the following form:
\be
\label{JHEP2}
 -\left(-\frac{12c\kappa^2}{l^2} + 1\right)\frac{1}{R}\sqrt{1 + \frac{R^2}{l^2}} = \eta\ ,
\ee
which has a solution if $\eta<- 1 + 12c\kappa^2/l^2$. The above choice of surface term
(junction condition) was motivated by AdS/CFT correspondence and well-defined 
HD variational procedure\cite{NOjhep}. Using it, the creation of dS and AdS brane universes 
in EGB gravity has been carefully studied in ref.\cite{NOjhep}
(also with the account of brane conformal anomaly\cite{zerbini}).
Nevertheless, misunderstanding of ambiguity of brane cosmology may lead 
to number of controversial claims.
For instance, in ref.\cite{AM}, the same problem of creation 
of dS and AdS branes in 5d EGB 
gravity as in ref.\cite{NOjhep} was re-addressed. 
In the notations of this paper, their junction condition is
\be
\label{GBc8}
\frac{1}{R}\sqrt{1 + \frac{R^2}{l^2}}
=\frac{l\bar\lambda}{2}\left[ 3 - \frac{4c\kappa^2}{l^2} + \left( 1 + \frac{4c\kappa^2}{l^2} \right)
\frac{l^2}{R^2}\right]^{-1}\ .
\ee
Here $\bar\lambda$ is brane tension.
Since $R_{(4)}=12/R^2$,  comparing (\ref{GBc8}) with (\ref{GBc6}), one trivially finds that (\ref{GBc8}) 
can be reproduced from (\ref{GBc6})  with the following choice for $f\left(R_{(0)}\right)$ 
\be
\label{GBc9}
f\left(R_{(0)}\right)= - \frac{8\zeta}{\xi} + \frac{4\zeta}{\xi^2}R_{(0)} 
+ \frac{4\zeta}{\xi^3}R_{(0)}^2\ln\frac{R_{(0)}}{\xi + R_{(0)}}\ .
\ee
Here  
$\zeta \equiv \left(2\bar\lambda/l\right)\left( 1 + 4c\kappa^2/l^2 \right)^{-1}\left( 1 - 12c\kappa^2/l^2 \right)^{-1}$ 
and $\xi \equiv \left(12/l^2\right)\left( 1 + 4c\kappa^2/l^2 \right)^{-1} \left( 3 - 4c\kappa^2/l^2 \right)$. 
Of course, one can suggest many more choices for $f$ and define ``new'' cosmologies with some of them 
being quite realistic.

To conclude, we demonstrated that brane cosmology is ambigious due to the function dependence 
of junction condition (or  due to the freedom in the choice of boundary action). This is general property 
of any (Einstein, EGB, etc.) brane-world gravity. There should exist fundamental physical principle 
(subject that brane approach is realistic) to fix this ambiguity and to make the brane cosmology 
to be predictable. So far, having in mind string/M-theory as a candidate for unified model it looks that 
only AdS/CFT related considerations may help for this purpose 
(but only partially as is seen in EGB theory).
It is remarkable that this point may be supported by using fundamental conformal  symmetry (also based 
on AdS/CFT) to fix the form of brane-world effective action\cite{conf}.

\noindent
{\bf Acknowledgments} 
This research has been supported in part by the Monbusho 
 of Japan under grant n.13135208 (S.N.), 
by grant 2003/09935-0 of FAPESP, Brazil (S.D.O.)
 and by project BFM2003-00620, Spain (S.D.O.).


\begin{thebibliography}{99}
\bibitem{RS1} L. Randall and R. Sundrum, {\sl Phys.Rev.Lett.} {\bf 83} (1999) 3370, hep-ph/9905221; 
 {\sl Phys.Rev.Lett.} {\bf 83} (1999) 4690, hep-th/9906064.
\bibitem{maartens} R. Maartens, {\sl Living Rev.Rel} {\bf 7} (2004) 1,
 gr-qc/0312059.
\bibitem{lidsey} J.E. Lidsey, {\sl Lect.Notes Phys.} {\bf 646} (2004) 357,
 astro-ph/0305528.
\bibitem{GH} G.W. Gibbons and S.W. Hawking, {\sl Phys.Rev.} {\bf D15} (1977) 2752.
\bibitem{Israel}  W. Israel, {\sl Nuovo Cim.} {\bf B44S10} (1966) 1, Erratum-ibid.{\bf B48} (1967) 463; 
{\sl Nuovo Cim.} {\bf B44} (1966) 1. 
\bibitem{CR} H.A. Chamblin and H.S. Reall, {\sl Nucl.Phys.} {\bf B562} (1999) 133, hep-th/9903225.
\bibitem{EJM} R. Emparan, C.V. Johnson and R.C. Myers, {\sl Phys.Rev.} {\bf D60} (1999) 104001, hep-th/9903238. 
\bibitem{zerbini} S. Nojiri, S.D. Odintsov and S. Zerbini, {\sl Phys.Rev.} {\bf D62} (2000) 064006, hep-th/0001192;
S.W. Hawking, T. Hertog and H.S. Reall, {\sl Phys.Rev.} {\bf D62} 043501, hep-th/0003052;
S. Nojiri and S.D. Odintsov, {\sl Phys.Lett.} {\bf B484} (2000)119, hep-th/0004097.
\bibitem{GB} S. Nojiri, S.D. Odintsov and S. Ogushi, {\sl Phys.Rev.} {\bf D65} (2002) 023521, hep-th/0108172; 
M. Cvetic, S. Nojiri and S.D. Odintsov, {\sl Nucl.Phys.} {\bf B628} (2002) 295, hep-th/0112045; 
J.E. Lidsey, S. Nojiri and S.D. Odintsov, {\sl JHEP} {\bf 0206} (2002) 026, hep-th/0202198.
\bibitem{GB1}
M. Giovannini, hep-th/0009172;
Y.M. Cho, I. Neupane and P.S. Wesson, hep-th/0104227;
B. Abdesselam and N. Mohammedi, hep-th/0110143;
C. Germani and C. Sopuerta, hep-th/0202060;
N. Mavromatos and J. Rizos, hep-th/0205299, hep-th/0008074;
J.E. Lidsey and N. Nunes, astro-ph/0303168;
N. Deruelle and J. Madore, gr-qc/0305004;
C. Barcelo, C. Germani and C. Sopuerta, gr-qc/0306072;
N. Deruelle and C. Germani, gr-qc/0306116;
G. Kofinas, R. Maartens and E. Papantonopoulos, hep-th/0307138;
B.C. Paul and M. Sami, hep-th/0312081;
H. Lee and G. Tasinato, hep-th/0401221;
M. Minamitsuji and M. Sasaki, hep-th/0404166;
M. Dehghani, hep-th/0312030, hep-th/0404118;
M. Sami and N. Dadhich, hep-th/0405016;
P. Wang and X. Meng, hep-th/0406170, hep-ph/0312113;
 O. Corradini, hep-th/0405038;
M. Sami, N. Savchenko and A. Toporensky, hep-th/0408140.
\bibitem{GB2}
S. Nojiri and S.D. Odintsov, hep-th/0109122;
E. Abdalla and L. Correa-Borbonet, hep-th/0109129;
R.G. Cai, hep-th/0109133, hep-th/0311240;
Y.M. Cho and I. Neupane, hep-th/0112227;
S. Nojiri, S.D. Odintsov and S. Ogushi, hep-th/0205187;
E. Gravanis and S. Willison, hep-th/0209076;
M. Dehghani, hep-th/0405206; 
M. Cvitan, S. Pallua and P. Prester, hep-th/0212029;
M. Fukuma, S. Matsuura and T. Sakai,hep-th/0212314;
I. Neupane, hep-th/0302132;
A. Padilla, gr-qc/0303082;
J.P. Gregory and A. Padilla, hep-th/0304250;
N. Deruelle, J. Katz and S. Ogushi, gr-qc/0310098;
R.G. Cai and Q.Guo, hep-th/0311020; M. Banados, hep-th/0310098;
G. Kofinas and E. Papantonopoulos, gr-qc/0401047;
T. Clunan, S. Ross and D. Smith, gr-qc/0402044;
S. Ogushi and M. Sasaki, hep-th/0407083;
\bibitem{NOjhep} S. Nojiri and S.D. Odintsov, {\sl JHEP} {\bf 0007} (2000) 049, hep-th/0006232.
\bibitem{AM} K.-s. Aoyanagi and K.-i. Maeda, hep-th/0408008
\bibitem{conf} S. Kanno and J. Soda, hep-th/0312106;
P. McFadden and N. Turok, hep-th/0409122.





\end{thebibliography}
\end{document}